\documentclass[onecolumn,preprint,showpacs,amsmath,amssymb,aps,pra,floatfix,nofootinbib]{revtex4}
\usepackage{graphicx,graphics,times}
\usepackage{bm}
\usepackage{longtable}

\def\be{\begin{equation}}
\def\ee{\end{equation}}
\def\ba{\begin{eqnarray}}
\def\ea{\end{eqnarray}}
\def\la{\langle}
\def\ra{\rangle}
\def\a{\alpha}

\def\m{\mu}
\def\n{\nu}
\def\h{\hskip 1cm}

\def\lo{\rightarrow}

\begin{document}
\vspace{1cm}
\title{Quantum Phase Transitions and Matrix Product States in Spin Ladders }
\vspace{3cm}
\author{M. Asoudeh}
\email{asoudeh@mehr.sharif.edu}\hspace{0.5cm}
\author{V. Karimipour\footnote{Corresponding Author}}
\email{vahid@sharif.edu}\hspace{0.5 cm}
\author{A. Sadrolashrafi}
\email{sadr_of_nobles@sharif.edu} \affiliation{Department of
Physics, Sharif University of Technology, P.O. Box 11365-9161,
Tehran, Iran} \vspace{2cm}
\begin{abstract}
We investigate quantum phase transitions in ladders of spin $1/2$
particles by engineering suitable matrix product states for these
ladders. We take into account both discrete and continuous
symmetries and provide general classes of such models. We also study
the behavior of entanglement between different neighboring sites
near the transition point and show that quantum phase transitions in
these systems are accompanied by divergences in derivatives of
entanglement.
\end{abstract}

\date{\today}
 \maketitle

\section{Introduction}\label{intro}
The basic paradigm of many body physics is to use analytical and
numerical tools to investigate the low-lying states and in
particular the ground state properties of a system governed by a
given Hamiltonian.  At very low temperatures, when thermal
fluctuations are dominated by quantum fluctuations, quantum phase
transitions can occur due to the change of character of the ground
state \cite{Sachdev}. What is exactly meant by "character" has been
investigated in numerous works \cite{Osterloh}, \cite{Nielsen},
\cite{VedralBose}, especially in recent years after the discovery of
exact measures \cite{Wootters} of entanglement(purely quantum
mechanical correlations). For example, it has been a matter of
debate whether a quantum phase transition is  always accompanied by
a divergence of
some property in entanglement of the ground state wave function. \\
Unfortunately, except for a few exactly solvable examples, the task
of finding the exact ground state of a given Hamiltonian is
notoriously difficult. As always in dealing with difficult problems,
one way round the difficulty is to investigate the inverse problem,
that is to start from states with pre-determined properties and
investigate quantum phase transitions which occur by smoothly
changing some continuous parameters of these states. The suitable
formalism for following this path is the matrix product formalism
\cite{MPSBasic1}, \cite{MPSBasic2}, \cite{WolfCirac, MPSrep} which
in recent years has been followed in constructing various models of
interacting spins
\cite{mpswork1,mpswork2,mpswork3,mpswork4,mpswork5,mpswork6}. In
this paper we want to take one step in this direction and in
particular, we want to construct as concrete models, ladders of spin
one-half particles and see what happens to the entanglement between
various spins when the system undergoes a phase transition. We
construct general class of models having a number of discrete and
continuous symmetries. These are symmetry under the exchange of legs
of the ladder, symmetry under spin flip, and symmetry under parity
(left-right reflection of the ladder) in addition to a continuous
symmetry, namely  rotation of spins around the $z$ axis or all three
axes (full rotational symmetry).\\
In these models we calculate the entanglement of one rung with the
rest of the ladder as measured by the von Neumann entropy of the
state of the rung and the entanglement of the two spins of a rung
with each other
as be measured by the concurrence of the same state.\\
We will see that in all these models quantum phase transitions occur
at a critical point of the coupling constant, and this point is
where a rung of the ladder becomes completely disentangled from the
rest of the ladder and the two spins of the rung become fully
entangled with each other. The derivatives of
these two types of entanglement are also divergent. \\
We should stress here that we are using the term phase transition in
a wider sense than usual \cite{WolfCirac}, that is we call any
discontinuity in an observable quantity (i.e. a two point
correlation function), a phase transition.\\

The structure of this paper is as follows. In section \ref{MPS}, we
review the formalism of Matrix Product States (MPS) with emphasis on
the symmetry properties of such states. In section \ref{Ladder}, we
specify the MPS construction to ladders of spin $1/2$ particles and
set the general ground for construction of concrete models. In this
same section we construct multi-parameter families of models which
have specific symmetries.  In section \ref{StateProperties}, we
study in detail the properties of the constructed states and in
particular calculate the exact correlation functions of spins on the
rungs. We also investigate the connection between quantum phase
transitions and divergence of entanglement properties in these
models. Finally we derive in the appendix, the Hamiltonian which
governs the interaction of these models for which the state we have
constructed is the exact ground state.
\section{Matrix Product States}\label{MPS}
First let us review the basics of matrix product states
\cite{MPSBasic2, WolfCirac}. Consider a homogeneous ring of $N$
sites, where each site describes a $d-$level state. The Hilbert
space of each site is spanned by the basis vectors $|i\ra, \ \
i=0,\cdots d-1$. A state
\begin{equation}\label{state}
    |\Psi\ra=\sum_{i_1,i_2,\cdots i_N}\psi_{i_1i_2\cdots
    i_N}|i_1,i_2,\cdots, i_N\ra
\end{equation}
is called a matrix product state if there exists $D$ dimensional
complex matrices  $A_i\in \mathbb{C}^{D\times D},\ \ i=0\cdots d-1$
such that
\begin{equation}\label{mat}
    \psi_{i_1,i_2,\cdots
    i_N}=\frac{1}{\sqrt{Z}}tr(A_{i_1}A_{i_2}\cdots A_{i_N}),
\end{equation}
where $Z$ is a normalization constant given by
\begin{equation}\label{z}
    Z=tr(E^N)
\end{equation}
and
\begin{equation}\label{E}
E:=\sum_{i=0}^{d-1} A_i^*\otimes A_i.
\end{equation}
Here we are restricting ourselves to translationally invariant
states, by taking the matrices to be site-independent. For open
boundary conditions, any state can be MPS-represented, provided that
we allow site-dependent matrices $A^{(k)}_i$, where $k$ denotes the
position of the site \cite{vidal, MPSrep}. The MPS representation
(\ref{mat}) is not unique and a transformation such as $A_i\lo UA_i
U^{-1}$ leaves the state invariant. In view of this we can find the
conditions on the matrices which impose discrete symmetries on the
state. The state $|\Psi\ra$ is reflection symmetric if there exists
a matrix $U$ such that  $A_i^T=UA_iU^{-1}$ where $A^T$ is the
transpose of $A$ and time-reversal invariant if there exist a matrix $V$ such that $A_i^*=VA_iV^{-1}$. \\
Let $O$ be any local operator acting on a single site. Then we can
obtain the one-point function on site $k$ of the chain $\la
\Psi|O(k)|\Psi\ra $ as follows:
\begin{equation}\label{1point}
    \la \Psi|O(k)|\Psi\ra = \frac{tr(E^{k-1}E_O E^{N-k})}{tr(E^N)},
\end{equation}
where
\begin{equation}\label{mpsop}
E_O:=\sum_{i,j=0}^{d-1}\la i|O|j\ra A_i^*\otimes A_j.
\end{equation}
In the thermodynamic limit $N\lo \infty$, equation (\ref{1point})
gives
\begin{equation}\label{1pointthermo}
    \la \Psi|O|\Psi\ra = \frac{\la
    \lambda_{max}|E_O|\lambda_{max}\ra}{\lambda_{max}},
\end{equation}
where we have used the translation invariance of the model and
$\lambda_{max}$ is the eigenvalue of $E$ with the largest absolute
value and $|\lambda_{max}\ra$ and $\la \lambda_{max}|$ are the right
and left eigenvectors corresponding to this eigenvalue, normalized
such that $\la \lambda_{max}|\lambda_{max}\ra=1$. Here we are
assuming that the largest eigenvalue of $E$ is non-degenerate.\\

The n-point functions can be obtained in a similar way. For example,
the two-point function $\la \Psi|O(k)O(l)|\Psi\ra$ can be obtained
as
\begin{equation}\label{2point}
\la \Psi|O(k)O(l)|\Psi\ra = \frac{tr(E_O(k)E_O(l)E^N)}{tr(E^N)}
\end{equation}
where $E_O(k):=E^{k-1}E_OE^{-k}$. Note that this is a formal
notation which allows us to write the n-point functions in a uniform
way, it does not require that $E$ is an invertible matrix. In the
thermodynamic limit the two point function turns out to be
\begin{equation}\label{2pointThermodynamicLimit}
\la \Psi|O(1)O(r)|\Psi\ra = \frac{1}{\lambda_{max}^{r}} {\sum_i
\lambda_i^{r-2} \la\lambda_{max}|E_{O}|\lambda_{i}\ra\la
\lambda_i|E_{O}|\lambda_{max}\ra}.
\end{equation}
For large distances $r\gg 1$, this formula reduces to
\begin{equation}\label{2pointrLarge}
\la \Psi|O(1)O(r)|\Psi\ra-\la \Psi|O|\Psi\ra^2
=\frac{\lambda_1^{r-2}}{\lambda_{max}^r} {\la
\lambda_{max}|E_{O}|\lambda_{1}\ra\la
\lambda_{1}|E_{O}|\lambda_{max}\ra},
\end{equation}
where $\lambda_1$ is the second largest eigenvalue of $E$ for which
the matrix element $\la \lambda_1|E_O|\lambda_{max}\ra$ is
non-vanishing and we have assumed that the eigenvectors of $E$ have
been normalized, i.e. $\la \lambda_i|\lambda_j\ra = \delta_{ij}$.
Thus the correlation length
 is given by
\begin{equation}\label{corr}
    \xi = \frac{1}{\ln \frac{\lambda_{max}}{\lambda_{1}}}.
\end{equation}

Any level crossing in the largest eigenvalue of the matrix $E$
signals a possible quantum phase transition. Also, due to
(\ref{corr}), any level crossing in the second largest eigenvalue of
$E$ implies the correlation length of the system has undergone a
discontinuous change. Here we are using a broader definition of
quantum phase transition, that is we call any non-analytical
behavior of a macroscopic property, a quantum phase transition
\cite{WolfCirac}. Of course in the models that we construct we
observe a more direct change of observable physical properties,
namely in one regime we have correlations between spin operators on
different sites and in the other we have no such correlation.

\subsection{Symmetries}
Consider now a local symmetry operator $R$ acting on a site as
$R|i\ra=R_{ji}|j\ra$ where summation convention is being used. $R$
is a $d$ dimensional unitary representation of the symmetry. A
global symmetry operator ${\cal R}:=R^{\otimes N}$ will then change
this state to another matrix product state
\begin{equation}\label{mpsPrime}
    \Psi_{i_1i_2\cdots i_N}\lo \Psi':=tr(A'_{i_1}A'_{i_2}\cdots
    A'_{i_N}),
\end{equation}
where
\begin{equation}\label{A'}
    A'_i:=R_{ij}A_j.
\end{equation}
A sufficient but not necessary condition for the state $|\Psi\ra$ to
be invariant under this symmetry is that there exist an operator
$U(R)$ such that
\begin{equation}\label{symm}
    R_{ij}A_j=U(R)A_iU^{-1}(R).
\end{equation}
Thus $R$ and $U(R)$ are two unitary representations of the symmetry,
respectively of dimensions $d$ and $D$. In case that $R$ is a
continuous symmetry with generators $T_a$, equation (\ref{symm}),
leads to
\begin{equation}\label{symmalg}
    (T_a)_{ij} A_j=[{\cal T}_a,A_i],
\end{equation}
where $T_a$ and ${\cal T}_a$ are the $d-$ and $D-$dimensional
representations of the Lie algebra of the symmetry. Equations
(\ref{symm}) and (\ref{symmalg}) will be our guiding lines in
defining states with prescribed symmetries.

\subsection{The Hamiltonian}
Given a matrix product state, the reduced density matrix of $k$
consecutive sites is given by
\begin{equation}\label{rhok}
    \rho_{i_1\cdots i_k, j_1\cdots j_k}=\frac{tr((A_{i_1}^*\cdots A_{i_k}^*\otimes A_{j_1}\cdots A_{j_k})E^{N-k})}{tr(E^N)}.
\end{equation}
The null space of this reduced density matrix includes the solutions
of the following system of equations
\begin{equation}\label{cc}
    \sum_{j_1,\cdots, j_k=0}^{d-1}c_{j_1\cdots
    j_k}A_{j_1}\cdots A_{j_k}=0.
\end{equation}
Given that the matrices $A_i$ are of size $D\times D$, there are
$D^2$ equations with $d^k$ unknowns. Since there can be at most
$D^2$ independent equations, there are at least $d^k-D^2$ solutions
for this system of equations. Thus for the density matrix of $k$
sites to have a null space it is sufficient that the following
inequality holds
\begin{equation}\label{dD}
    d^k\ >\ D^2.
\end{equation}
Let the null space of the reduced density matrix be spanned by the
orthogonal vectors $|e_{\a}\ra, \ \ \ (\a=1, \cdots  s,\geq
d^k-D^2)$. Then we can construct the local hamiltonian acting on $k$
consecutive sites as
\begin{equation}\label{h}
    h:=\sum_{\a=1}^s \mu_{\a} |e_{\a}\ra\la e_{\a}|,
\end{equation}
where $\mu_{\a}$'s are positive constants. These parameters together
with the parameters of the vectors $|e_i\ra $ inherited from those
of the original matrices $A_i$, determine the total number of
coupling constants of the Hamiltonian.  If we call the embedding of
this local Hamiltonian into the sites $l$ to $l+k$ by $h_{l,l+k}$
then the full Hamiltonian on the chain is written as
\begin{equation}\label{H}
    H=\sum_{l=1}^N h_{l,l+k}.
\end{equation}
The state $|\Psi\ra$ is then a ground state of this hamiltonian with
vanishing energy. The reason is as follows:
\begin{equation}\label{Hrho}
\la \Psi|H|\Psi\ra=tr(H|\Psi\ra\la\Psi|)=\sum_{l=1}^N
tr(h_{l,l+k}\rho_{l,l+k})=0,
\end{equation}
where $\rho_{l,k+l}$ is the reduced density matrix of sites $l$ to
$l+k$ and in the last line we have used the fact that $h$ is
constructed from the null eigenvectors of $\rho$ for $k$ consecutive
sites. Given that $H$ is a positive operator, this proves the
assertion.

\section{Ladders of spin one-half particles}\label{Ladder}
We now specify the above generalities to a ladder of spin $1/2$
particles. The ladder consists of $N$ rungs and obeys periodic
boundary conditions. To each rung of the ladder we associate four
matrices $A_{00},\ A_{01},\ A_{10}$ and $A_{11}$ respectively
pertaining to the local states $|0,0\ra,\ |0,1\ra, \ |1,0\ra $ and
$|1,1\ra$. Here we are using the qubit notation which corresponds to
the spin notation as $|0\ra=|+\ra$ and $|1\ra=|-\ra.$ The first and
the second indices refer respectively to the states of sites on the
legs 1 and 2 as shown in figure (1). The rungs are numbered from 1
to N with a periodic boundary condition. The operator
$\sigma^{(i)}_{ak}$ refers to the Pauli operator $\sigma_a$ on leg
number $i$ and the rung number $k$. The total spin operator on a
rung at site $k$ is denoted by ${\bf
S}_k:=\frac{1}{2}(\pmb{\sigma}_k^{(1)}+\pmb{\sigma}_k^{(2)})$.

\subsection{MPS operators for observables}
Let us label the legs of the ladder by $1$ and $2$ as in figure (1).
In each $A_{ij}$ the first and the second indices refer respectively
to legs number 1 and 2.
\begin{figure}
\centering
    \includegraphics[width=8cm,height=2.5cm,angle=0]{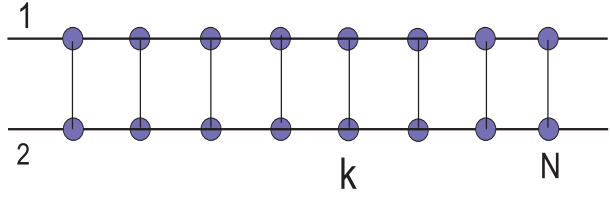}
    \caption{(Color Online) A spin 1/2 ladder, an operator ${\sigma}_{ak}^{(i)}$ refers to the Pauli operator $\sigma_a$ on rung $k$ and leg number $(i)$. }
\end{figure}\label{ladder}
Any operator corresponding to leg number $i$  is designated with a
superscript $(i)$.  With these conventions, one can easily use
(\ref{mpsop}) and write down the MPS operator corresponding to an
observable. For example, for the magnetization in the $x$ and $z$
directions in rung 1, we have respectively
\begin{equation}\label{Ex}
    E_{\sigma_x^{(1)}}=A_{00}\otimes A_{10}+A_{10}\otimes A_{00}+A_{01}\otimes A_{11}+A_{11}\otimes A_{01},
\end{equation}
and
\begin{equation}\label{Ez}
    E_{\sigma_z^{(1)}}=A_{00}\otimes A_{00}+A_{01}\otimes A_{01}-A_{10}\otimes A_{10}-A_{11}\otimes A_{11}.
\end{equation}
For the total magnetization in the $z$ direction in a rung we have
\begin{equation}\label{Eztotal}
 E_{S_z}=\frac{1}{2}(E_{\sigma_z^{(1)}}+E_{\sigma_z^{(2)}})=(A_{00}\otimes A_{00}-A_{11}\otimes A_{11}).
\end{equation}
Other MPS operators can be obtained in a similar way.

\subsection{Symmetries of spin ladders} In this paper we restrict the dimensions of our
matrices to $D=2$, and demand that our models have time reversal
symmetry so that the matrices $A_{ij}$ are chosen to be real. We
will be looking for models which have rotational symmetry in the
$x-y$ plane of spin space. Thus we require that there be a matrix
${\cal T}_z$ such that
\begin{equation}\label{Tz}
    [{\cal T}_z, A_{00}]=A_{00}, \ \ \ \ [{\cal T}_z, A_{11}]=-A_{11}, \ \ \
    \ [{\cal T}_z, A_{01}]=[{\cal T}_z, A_{10}]=0.
\end{equation}
It is an easy exercise to show that the solution of these equations
is
\begin{equation}\label{TzSolution}
A_{01}=\left(\begin{array}{cc} a & 0 \\ 0 & b \end{array}\right), \
\ \ \ A_{10}=\left(\begin{array}{cc} a' & 0 \\ 0 & b'
\end{array}\right), \ \ \ \ A_{00}=\left(\begin{array}{cc} 0 & g \\ 0 & 0
\end{array}\right), \ \ \ \ A_{11}=\left(\begin{array}{cc} 0 & 0 \\ h & 0
\end{array}\right),
\end{equation}
where ${\cal T}$ is found to be ${\cal
T}_z=\frac{1}{2}\left(\begin{array}{cc} 1 & 0
\\ 0 & -1 \end{array}\right)$ and we have excluded the solutions with $g=0$
or $h=0$, which lead to trivial uncorrelated states. Hereafter we
use the freedom in re-scaling the matrices (without changing the
matrix product state) to set the parameter $h=1$. Let us now consider extra discrete symmetries in addition to the above continuous symmetry.
These are as follows:\\

{\textbf{a: Spin flip symmetry}} represented by a matrix $X$, such
that
\begin{equation}\label{X}
    XA_{i,j}X^{-1}=\epsilon A_{\overline{i},\overline{j}},
\end{equation}
where $\epsilon = \pm 1$ and $\overline{i}$ means $i+1\ \ mod \ \
2$. This symmetry imposes the following condition on the solution
(\ref{TzSolution})
\begin{equation}\label{Xsolution}
    a'=\epsilon b, \h b'=\epsilon a,
\end{equation}
where $X$ is found to be $X=\left(\begin{array}{cc} 0 & g \\
\epsilon & 0
\end{array}\right)$.

{\textbf{b: Symmetry under the exchange of legs of the ladder}}
represented by a matrix $Y$ such that
\begin{equation}\label{Y}
    YA_{i,j}Y^{-1}=\eta A_{j,i},
\end{equation}
where $\eta = \pm 1$. This symmetry imposes the following condition
on (\ref{TzSolution})
\begin{equation}\label{Ysolution}
    a'=\eta a, \h b'=\eta b,
\end{equation}
where $Y$ is found to be $Y=\left(\begin{array}{cc} 1 & 0 \\ 0 &
\eta
\end{array}\right)$.

{\textbf{c: Parity}} or a left-right symmetry, represented by a
matrix $\Pi$, such that
\begin{equation}\label{pi}
 \Pi A_{ij}\Pi^{-1} =\sigma A^T_{ij},
\end{equation}
with $\sigma=\pm 1$. \\
This symmetry imposes the condition
\begin{equation}\label{Pisolution}
    b=\sigma a, \h b'=\sigma a',
\end{equation}
on (\ref{TzSolution}) where $\Pi$ is found to be $\Pi=\left(\begin{array}{cc} 0 & 1 \\
\sigma &
0\end{array}\right)$.\\
Thus when we impose any of these discrete symmetries we are left
with two three-parameter families of models, each family being
distinguished by the discrete parameter $\epsilon$, $\eta$ or
$\sigma$. \\
It is now readily seen that imposing any two of these symmetries
makes the model symmetric under the third one too. A model which has
all three symmetries is defined by the following set of matrices:

\begin{equation}\label{Z23Solution}
A_{01}=\left(\begin{array}{cc} a & 0 \\ 0 & \sigma a
\end{array}\right), \ \ \ \ A_{10}=\left(\begin{array}{cc} \epsilon
\sigma a & 0 \\ 0 & \epsilon a
\end{array}\right), \ \ \ \ A_{00}=\left(\begin{array}{cc} 0 & g \\ 0 & 0
\end{array}\right), \ \ \ \ A_{11}=\left(\begin{array}{cc} 0 & 0 \\ 1 & 0
\end{array}\right).
\end{equation}
Such matrices satisfy (\ref{Y}) with $\eta=\epsilon\sigma$. Thus
equation \ref{Z23Solution} defines four two-parameter families of
models on spin ladders which have $SO(2)$ symmetry in addition to
three types of $Z_2$ symmetries. The families are distinguished by
the
pair of discrete parameters $(\epsilon,\sigma)$. \\

\textbf{d: Full rotational symmetry} Let us now see if we can
construct models which have full $SU(2)$ symmetry, symmetry under
rotations in the spin space. To this order we note that to have full
rotational symmetry, the matrices defined by
\begin{equation}\label{red}
    B_{1,1}:=A_{00},\ \ \ \
    B_{1,0}:=\frac{1}{\sqrt{2}}(A_{01}+A_{10}),\ \ \ \
    B_{1,-1}:=A_{11},
\end{equation}
and
\begin{equation}\label{redSpin0}
    B_{0,0}:=\frac{1}{\sqrt{2}}(A_{01}-A_{10}),
\end{equation}
should respectively transform like the spin $1$ and spin $0$
representations of the $su(2)$ algebra, that is, we should have
\begin{equation}\label{TT}
   [{\cal T}_a, B_{0,0}]=0, \h a=x,\ y,\ z,
\end{equation}
and
\begin{equation}\label{TT2}
   [{\cal T}_z, B_{l,m}]=m \ B_{l,m}, \h
   [{\cal T}_{\pm}, B_{l,m}]= \sqrt{2-m(m\pm 1)}B_{l,m\pm 1}, \h
   l=1, \ m=-1,0,1,
\end{equation}
where ${\cal T}_a=\frac{1}{2}\sigma_a, \  a=x,y,z$ form the two
dimensional representation of the $su(2)$ algebra. It is well known
that the matrices $(\sigma_-, \frac{1}{\sqrt{2}}\sigma_z,
-\sigma_+)$ transform like a vector under the adjoint action of
$SU(2)$. The matrix $B_{0,0}$ should also be a multiple of identity.
Thus we should set $g=-1$, and satisfy the following equations
\begin{equation}\label{uuu}
    A_{01}+A_{10}=\sigma_z, \h A_{01}-A_{10}=uI.
\end{equation}
This puts the constraints
\begin{equation}\label{constraint}
    a+a'=1, \ \ \ b+b'=-1,\ \ \ a-a'=u, \ \ \ b-b'=u,
\end{equation}
with the unique solution
\begin{equation}\label{solU}
    a=\frac{u+1}{2}, \ \ \ \ a'=\frac{1-u}{2},\ \ \ \
    b=\frac{u-1}{2}, \ \ \ \ b'=\frac{-1-u}{2},
\end{equation}
where $u$ is an arbitrary real parameter. This will give us a
one-parameter family of models with full rotational symmetry and
spin-flip symmetry
\begin{equation}\label{RotSolution}
A_{01}=\left(\begin{array}{cc} \frac{u+1}{2} & 0 \\ 0 &
\frac{u-1}{2}
\end{array}\right), \ \ \ \ A_{10}=\left(\begin{array}{cc} \frac{1-u}{2} & 0 \\ 0 & -\frac{u+1}{2}
\end{array}\right), \ \ \ \ A_{00}=\left(\begin{array}{cc} 0 & -1 \\ 0 & 0
\end{array}\right), \ \ \ \ A_{11}=\left(\begin{array}{cc} 0 & 0 \\ 1 & 0
\end{array}\right).
\end{equation}

\textbf{Remark:}  Note that comparison of these parameters with the
constraints (\ref{Xsolution}, \ref{Ysolution}) and
(\ref{PiSolution}) shows that full rotational symmetry is compatible
with spin-flip symmetry for arbitrary values of the parameter $u$
and compatible with parity or leg-exchange symmetries only for
$u=0$.\\

The model (\ref{RotSolution}) has already been studied in
\cite{PolishPaper}. To see the correspondence with that work, we can
collect the above matrices in a vector-valued matrix ${\cal A}$
defined as ${\cal A}=\sum_{i,j}A_{ij}|i,j\ra$. In view of our
notation for spins $(|0\ra=|+\ra, \ |1\ra=|-\ra)$ and the notation
of \cite{PolishPaper} in which the single and the triplet states are
denoted respectively by $|s\ra$ and $(|t_1\ra, \ |t_0\ra, \
|t_{-1}\ra $ the matrix ${\cal A} $ derived from (\ref{RotSolution})
becomes
\begin{equation}\label{PolishA}
{\cal A}=\frac{1}{\sqrt{2}}\left(\begin{array}{cc} u|s\ra + |t_0\ra
& -\sqrt{2} |t_1\ra
\\ \sqrt{2}|t_{-1}\ra & u|s\ra - |t_0\ra
\end{array}\right),
\end{equation}
which modulo an overall constant is identical to the matrix given in
\cite{PolishPaper}.

\section{Properties of the states and correlation functions}\label{StateProperties} We now
study the properties of the states constructed above. For simplicity
we consider in detail only two general classes. The first class is
defined by (\ref{Z23Solution}) and has $SO(2)$ symmetry in addition
to three $Z_2$ symmetries, and the second class is defined by
(\ref{RotSolution}) which has full rotational symmetry in addition
to one $Z_2$ symmetry, the spin flip symmetry. As mentioned in the
previous section, full rotational symmetry is compatible only with
spin flip symmetry for generic values of the parameter $u$ and is
compatible with the other two symmetries only when $u=0$. In order
to derive results which can be specialized to the two classes
mentioned above we study in detail the properties of the state,
defined by the equation (\ref{Xsolution}). The matrices are now
given by
\begin{equation}\label{Properties}
A_{01}=\left(\begin{array}{cc} a & 0 \\ 0 & b
\end{array}\right), \ \ \ \ A_{10}=\left(\begin{array}{cc} \epsilon b & 0 \\ 0 & \epsilon a
\end{array}\right), \ \ \ \ A_{00}=\left(\begin{array}{cc} 0 & g \\ 0 & 0
\end{array}\right), \ \ \ \ A_{11}=\left(\begin{array}{cc} 0 & 0 \\ 1 & 0
\end{array}\right).
\end{equation}
The matrix $E$ for this class has the following form
\begin{equation}\label{ESolution}
    E=\left(\begin{array}{cccc} a^2+b^2 & 0 & 0 &g^2 \\ 0 & 2ab & 0 & 0 \\ 0 & 0 & 2ab & 0 \\ 1 & 0 &
    0 &
    a^2+b^2\end{array}\right),
\end{equation}
with eigenvalues
$$\lambda_1=a^2+b^2+g,\ \ \ \ \ \lambda_2=a^2+b^2-g,\ \ \ \ \
\lambda_3=\lambda_4=2ab.$$ For $g>0$, the largest eigenvalue is
$\lambda_1 $ and for $g<0$ it is $\lambda_2$. Hence the point $g=0$
is a point of phase transition. The right and left eigenvectors of
$E$ are simply obtained and one can determine all the relevant
quantities of the ground state in closed form, in straightforward
way after some rather lengthy calculations. The reduced one-rung
density matrix is obtained from (\ref{rho}):
\begin{equation}\label{RhoSolution}
    \rho=\frac{1}{Q}\left(\begin{array}{cccc} |g| & 0 & 0 & 0 \\
0 & a^2+b^2 & 2\epsilon a b & 0 \\ 0 & 2\epsilon ab & a^2 + b^2 & 0
\\ 0 & 0 & 0 & |g|
    \end{array}\right)
\end{equation}
where $Q:=2a^2+2b^2+2|g|$. This matrix can be rewritten as
\begin{equation}\label{RhoSolutionKetBra}
    \rho =\frac{1}{Q}\left[|g|(|00\ra\la 00|+|11\ra\la
    11|) + 2\epsilon ab
(|01\ra\la 10|+|10\ra\la 01|)+(a^2+b^2)(|01\ra\la 01|+|10\ra\la
10|)\right],
\end{equation}
or
\begin{equation}\label{RhoSolutionKetBra2}
    \rho =\frac{1}{Q}\left[|g|(|t_1\ra\la t_1|+|t_{-1}\ra\la
    t_{-1}|) + \frac{1}{2}(a+\epsilon b)^2|t_0\ra\la
    t_0|+\frac{1}{2}(a-\epsilon b)^2|s\ra\la s|\right],
\end{equation}
where we have used the notation introduced in equation
(\ref{PolishA}).\\

From this density matrix one can obtain a lot of information about
the observables pertaining to a single rung. For example it is
readily seen that the average magnetization at each single site and
hence each single rung is zero, i.e.
\begin{equation}\label{magnetization}
   \la  \pmb{\sigma}^{(1)}\ra=\la \pmb{\sigma}^{(2)}\ra=0,\h \la {\bf S}\ra=0,
\end{equation}
where ${\bf S}$ is the spin of a rung, implying an
anti-ferromagnetic state in which every single site is in a
completely mixed states. It is also seen that
\begin{eqnarray}\label{CorrelationSolution}
&& \la \sigma_z^{(1)}\sigma_{z}^{(2)}\ra =
\frac{|g|-a^2-b^2}{|g|+a^2+b^2},\h \la \sigma_{{\bf
n}}^{(1)}\sigma_{{\bf n}}^{(2)}\ra = \frac{2\epsilon a
b}{|g|+a^2+b^2},\ \ \
\end{eqnarray}
where ${\bf n}$ is any unit vector in the $x-y$ plane.\\
Defining the total spin of a single rung as ${\bf
S}:=\frac{1}{2}(\pmb{\sigma}^{(1)}+\pmb\sigma^{(2)})$, we find from
the above result that
\begin{equation}\label{S2Solution}
    \la {\bf S}^2\ra =
    \frac{(a+\epsilon b)^2+2|g|}{a^2+b^2+|g|}.
\end{equation}
Thus for $|g|\lo \infty$, each rung will be in a mixture of spin one
states, but for $g=0$, the spin one and spin zero multiplets can
mix, depending on the value of $a$ and $b$ and $\epsilon$. The
entanglement of this rung with the rest of the lattice is measured
by the von-Neumann entropy of this state, defined as $S=-tr(\rho\log
\rho)$.

From the eigenvalues of the one-rung density matrix,
\begin{equation}\label{Eigenvalues of Rho}
    \a_1=\a_2 = \frac{|g|}{Q},\ \ \ \a_{3}=\frac{(a-b)^2}{Q},\
    \ \ \a_4=\frac{(a+b)^2}{Q}.
\end{equation}
we readily find
\begin{equation}\label{EntropySolution}
    S=\log Q-\frac{1}{Q}\left[|g|\log g^2 + (a+b)^2 \log (a+b)^2 + (a-b)^2 \log
    (a-b)^2\right].
\end{equation}
As $|g|\lo \infty$ $S\lo \log 2 = 1 $. Therefore in this limit each
rung is still entangled with the rest of the ladder. This means that
a rung is not in a pure state and the state of the ladder is not a
product of single rung states. \\  The entanglement of the two spins
of a single rung with each other is measured by the concurrence
\cite{Wootters} of this state which for the density matrix
(\ref{RhoSolution}) is given by
\begin{equation}\label{concurrence2}
    C=max(0,2\a_{max}-1),
\end{equation}
where $\a_{max}$ is the largest eigenvalue of the matrix $\rho$. A
careful analysis of the eigenvalues shows that
\begin{equation}\label{concurrence}
    C= max (0, \frac{2|ab|-|g|}{a^2+b^2+|g|}).
    \end{equation}
As $|g|\lo \infty$, $C\lo 0$ which means that although each single
rung is not a pure state, it is a separable state. In fact these can
also be verified directly by looking at the one-rung density matrix
in this limit. From (\ref{RhoSolutionKetBra}) we have
\begin{equation}\label{rhoginfinity}
    \rho(g\lo\infty)=\frac{1}{2}(|00\ra\la 00|+|11\ra\la 11|).
\end{equation}
 Finally we
calculate the correlation functions of different components of spins
of the rung as a function of the distance between the rungs. We find
from (\ref{2pointThermodynamicLimit}) that
\begin{equation}\label{CorrelationZSolution}
    \la S_{z,1}S_{z,r}\ra= - g^2
    \frac{(a^2+b^2-|g|)^{r-2}}{(a^2+b^2+|g|)^r}=\la S_{z,1}S_{z,2}\ra e^{-(r-2)/\xi_z},
\end{equation}
with a longitudinal correlation length
\begin{equation}\label{LongitudinalLength}
    \xi_{z}:=\frac{1}{\ln(\frac{a^2+b^2+|g|}{|a^2+b^2-|g||})},
\end{equation}
and
\begin{equation}\label{CorrelationXSolution}
    \la S_{{\bf n},1}S_{{\bf n},r}\ra= (sgn(g)+\epsilon)(a+sgn(g)b)^2\frac{|g|}{2}
    \frac{(2ab)^{r-2}}{(a^2+b^2+|g|)^r}=\la S_{{\bf n},1}S_{{\bf
    n},2}\ra e^{-(r-2)/\xi_{{\bf n}}}
\end{equation}
with a transverse correlation length
\begin{equation}\label{TransverseLength}
    \xi_{{\bf n}}:=\frac{1}{\ln(\frac{a^2+b^2+|g|}{|2ab|})}.
\end{equation}
It is seen that the longitudinal correlation length depends on a
single parameter $x:=\frac{g}{a^2+b^2}$ as
$\xi_z^{-1}=\ln(\frac{1+|x|}{1-|x|})$ and the transverse correlation
length depends on $x$ and another parameter
$\mu:=\frac{2|a||b|}{a^2+b^2}\leq 1$ as $\xi_{{\bf
n}}^{-1}=\ln(\frac{1+|x|}{\mu})$. Figure (2) shows the behavior of
these correlation functions for different values of the parameters
$x$ and $\mu$.

\begin{figure}\label{correlationLengths}
\centering
    \includegraphics[width=8cm,height=6cm,angle=0]{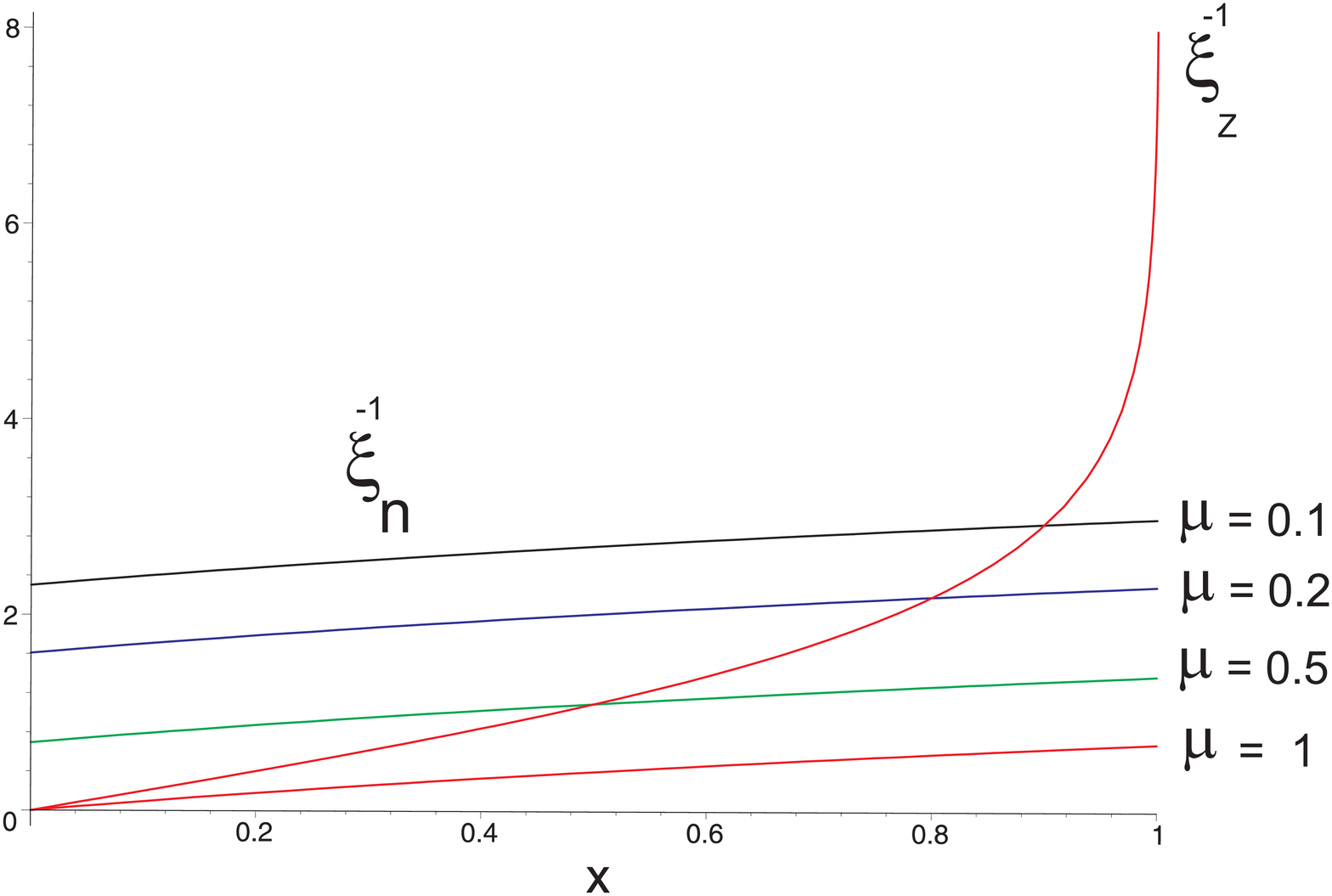}
    \caption{(Color Online) The longitudinal and transverse correlation lengths of matrix product states \ref{Properties},
     as a function of the parameter $x$
     for different values of the parameter $\mu:=\frac{2|ab|}{a^2+b^2}$.  At $\mu=1$
    both the transverse and longitudinal correlation lengths diverge at $x=0$.}
\end{figure}
An interesting point about transverse correlation functions is that
depending on the sign of $\epsilon$, it becomes identically zero, on
one side of the $g-$ axis and different from zero on the other side.
Thus if one insists on a definition of an order parameter to be zero
in one phase and non-zero in the other, then we can safely say that
in these models, the transverse correlation function is an order
parameter which signals a quantum phase transition. \\

Let us see study some limiting cases in these general models. At the
point of phase transition $g=0$, as seen from
(\ref{RhoSolutionKetBra}), the state of a rung becomes a mixture of
spin-zero states, and the two spins of a rung become fully
anti-correlated in the $z$ direction, a fact which is reflected in
(\ref{CorrelationSolution}). Near this point the correlation length
$\xi_z$ becomes very large as seen from (\ref{LongitudinalLength}),
although the amplitude becomes small since it is proportional to
$|g|$. Thus at $g=0$ there is no long-range order in the model. Also
from (\ref{concurrence}) and (\ref{Eigenvalues of Rho}) it is
readily seen that at $g=0$, the concurrence (or entanglement) of the
two spins of a rung becomes maximum and equal to
\begin{equation}\label{conmax}
    C(g=0)=\frac{2|ab|}{a^2+b^2}.
\end{equation}
On the other hand from (\ref{EntropySolution}) we see that as we
approach the point $g=0$ from both sides, the von-Neumann entropy
decreases. In addition, the derivative of both types of entanglement
become singular at $g=0$. These facts show that the point of phase
transition in these systems, is a point where the spins of a rung
become highly entangled with each other and each rung becomes only
slightly entangled to the rest of the lattice. \\
We can also obtain the explicit form of the state in this limit.
From (\ref{Properties}), and (\ref{mat}) we see that in this limit,
no two spins in a rung can be in a $\left(\begin{array}{c} 0
\\ 0 \end{array}\right)$ state. A little reflection shows that they can not be in the state $\left(\begin{array}{c} 1
\\ 1 \end{array}\right)$ either, since in the absence of $A_{00}$, the only string of matrices with non-zero trace is a string of matrices $A_{01}$ and $A_{10}$ in arbitrary order. Since these matrices commute
with each other, the resulting state has a simple description.
Define two local states of a rung as

\begin{equation}\label{ud}
|u\ra:=\left(\begin{array}{c} 0
\\ 1 \end{array}\right)\equiv \left(\begin{array}{c} +
\\ - \end{array}\right),\h  |d\ra:=\left(\begin{array}{c} 1
\\ 0 \end{array}\right)\equiv \left(\begin{array}{c} -
\\ + \end{array}\right),
\end{equation}
where for convenience we have re-introduced the spin notations $+$
and $-$ instead of qubit notation $0$ and $1$. Define a global
un-normalized state $|u^k,d^{N-k}\ra$ to be the equally weighted
linear combination of all states with $k$ local $u$ states and $N-k$
local $d$ state. Then from (\ref{Properties}) and (\ref{mat}) we
find that the ground state of the chain in the limit $g=0$ is given
by
\begin{equation}\label{Psig=0}
    |\Psi\ra = \frac{1}{\sqrt{Z}}\sum_{k=0}^{N} \left[ a^k(\epsilon b)^{N-k}+b^k(\epsilon
    a)^{N-k}\right]|u^k,d^{N-k}\ra,
\end{equation}
where $Z$ is the normalization constant given by
\begin{equation}\label{Zg=0}
    Z = \sum_{k=0}^N\left(\begin{array}{c} N
\\ k \end{array}\right)\left[a^k(\epsilon b)^{N-k}+b^k(\epsilon
a)^{N-k}\right]^2 = 2\left[(a^2+b^2)^N+(2ab)^N\right].
\end{equation}
At the other extreme when $|g|\lo \infty $, the state of each rung
becomes a mixture of fully aligned spins either in the positive or
negative $z$ direction. This is also reflected in
(\ref{CorrelationSolution}). In this limit a rung becomes entangled
with the rest of the lattice, sine $\lim_{|g|\lo \infty}S(\rho)=1$
and the two spins of a rung become disentangled
from each other since $C=0$.\\
The explicit form of the state can also be obtained in this limit.
In this case the only strings of matrices with non-vanishing traces
are strings of $A_{00}$ and $A_{11}$ in alternating order. Thus if
we define two local states

\begin{equation}\label{plusminus}
|t_1\ra:=\left(\begin{array}{c} 0
\\ 0 \end{array}\right)\equiv \left(\begin{array}{c} +
\\ + \end{array}\right),\h  |t_{-1}\ra:=\left(\begin{array}{c} 1
\\ 1 \end{array}\right)\equiv \left(\begin{array}{c} -
\\ - \end{array}\right),
\end{equation}
then the ground state in this limit will be a GHZ state
\begin{equation}\label{PsiG}
    |\Psi\ra = \frac{1}{\sqrt{2}}(|t_1,t_1,\cdots, t_1,t_1\ra
    +|t_{-1},t_{-1},\cdots,t_{-1},t_{-1}\ra).
\end{equation}
In the next subsection we specialize these results to the two
classes we discussed in the beginning of this subsection.

\subsection{Class A}\label{Examples} This is the class which has the SO(2) symmetry
(rotation around the $z$ axis in spin space) and three $Z_2$
symmetries (spin flip, parity, leg exchange). For this class we have
from (\ref{Pisolution}) that $b=\sigma a$. The parameter $x$ defined
after equation (\ref{TransverseLength}) becomes equal to
\begin{equation}\label{x}
    x:=\frac{g}{2a^2}.
\end{equation}
Insertion of $b=\sigma a$ in various quantities of the previous
subsection shows that all the quantities can be expressed as a
function of $x$, namely we find
\begin{equation}\label{modelA}
    \la \sigma_z^{(1)}\sigma_z^{(2)}\ra = \frac{|x|-1}{|x|+1}, \h  \la \sigma_{\bf n}^{(1)}\sigma_{\bf n}^{(2)}\ra =
    \frac{\epsilon\sigma}{|x|+1},
\end{equation}
\begin{equation}\label{ModelACorrelationsZ}
\la S_{z,1}S_{z,r}\ra = -x^2 \frac{(1-|x|)^{r-2}}{(1+|x|)^r}, \h
\xi_z = \frac{1}{\ln(\frac{1+|x|}{1-|x|})},
\end{equation}
and
\begin{equation}\label{ModelACorrelationsX}
\la S_{{\bf n},1}S_{{\bf n},r}\ra = (1+\epsilon\  sgn(x))(1+\sigma\
sgn(x)) |x| \frac{\sigma^r}{(1+|x|)^r}, \h \xi_{\bf n} =
\frac{1}{\ln(1+|x|)}.
\end{equation}

We also find
\begin{equation}\label{SmodelA}
    S = \frac{|x|}{1+|x|}(1-\log |x|)+\log(1+|x|),
\end{equation}
and

\begin{equation}\label{concurrenceA}
    C= max(0, \frac{1-|x|}{1+|x|}).
    \end{equation}
Figure (2) show the entropy and the concurrence of the state of a
single rung for this model.
\begin{figure}\label{FinalLadderCurvesABnew}
\centering
    \includegraphics[width=8cm,height=6cm,angle=0]{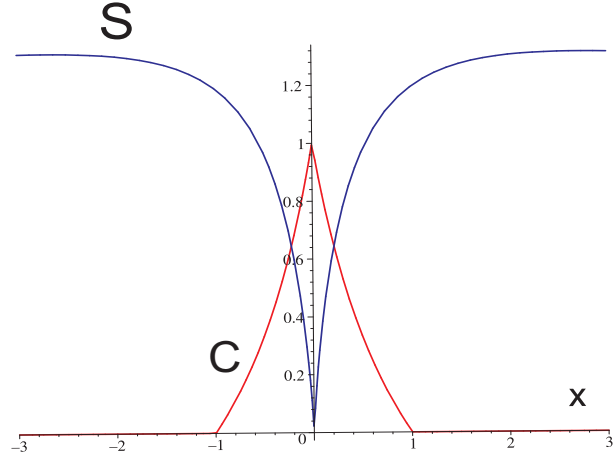}
    \caption{(Color Online) The von Neumann entropy of one single rung and its concurrence in models {\bf A}, as functions of the parameter $x:=\frac{g}{2a^2}$, which is taken to be dimensionless.}
\end{figure}

\subsection{Class B}\label{ExamplesRot} This is the class which has full SO(3) symmetry
(rotation in spin space) and one discrete symmetry (spin flip) for
generic values of $u$ and all the three $Z_2$ symmetries for $u=0$.
For this class we have from (\ref{RotSolution}) $g=\epsilon=-1$,
$a=\frac{u+1}{2}$ and $b=\frac{u-1}{2}$. Inserting these values in
the equations of the previous section we find the following:

\begin{equation}\label{modelB}
    \la \sigma_{\bf n}^{(1)}\sigma_{\bf n}^{(2)}\ra = \frac{1-u^2}{3+u^2},
\end{equation}
and
\begin{equation}\label{modelBcorrelations}
\la S_{{\bf n},1}S_{{\bf n},r}\ra = -4
\frac{(u^2-1)^{r-2}}{(u^2+3)^r},
\end{equation}
where ${\bf n}$ is any direction in the spin space. The eigenvalues
of the one-rung density matrix (\ref{Eigenvalues of Rho}) in this
case will be
\begin{equation}\label{Eigenvalues of Rho rot}
    \a_1=\a_2 = \a_{3}=\frac{1}{u^2+3},\
    \ \ \a_4=\frac{u^2}{u^2+3}.
\end{equation}
Thus we find
\begin{equation}\label{SmodelB}
    S = \log (u^2+3)- \frac{u^2\log u^2}{u^2+3},
\end{equation}
and
\begin{equation}\label{concurrenceB}
    C= max (0, \frac{u^2-3}{u^2+3}).
    \end{equation}
Figure (3) show the entropy and the concurrence of the state of a
single rung for this model.
\begin{figure}\label{Ucurves}
\centering
    \includegraphics[width=8cm,height=6cm,angle=0]{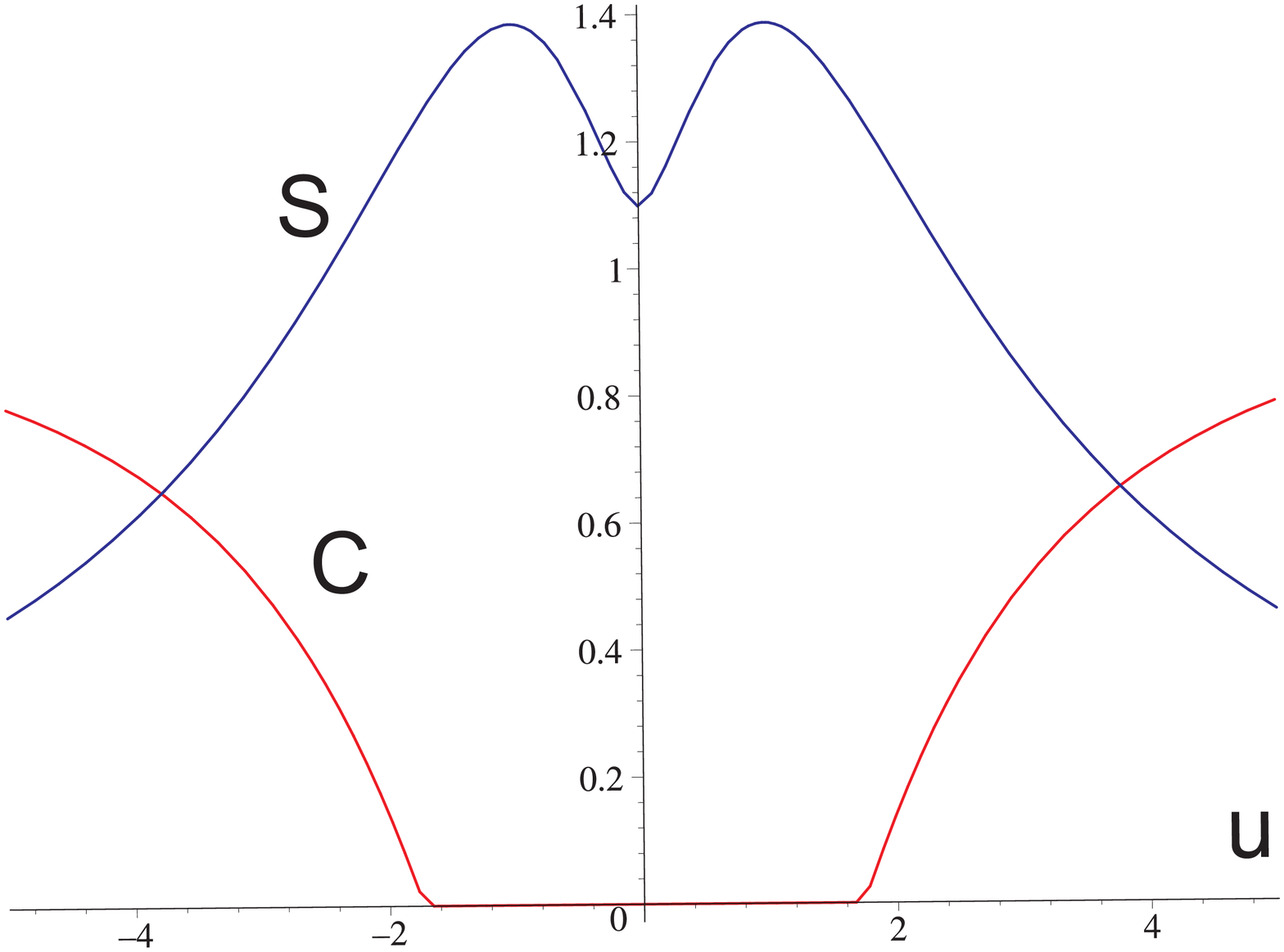}
    \caption{(Color Online)The von Neumann entropy of one single rung and its concurrence in model {\bf B}
    as functions of the parameter $u$, which is taken to be dimensionless.}
\end{figure}

\section{Discussion}
Although ladders of spin 3/2 models has already been studied in the
context of vertex state models in \cite{mpswork3}, in which the
ground state is constructed by a suitable concatenation of vertices
assigned to single sites, the method developed in this work, in
which each single rung is considered as a single site in a
hyper-chain and the ground state is constructed as a  matrix product
state seems to us as more powerful and very easy to generalize to
other spin models. In this work we have applied the matrix product
formalism for construction of models on spin $1/2$ ladders. These
models have been constructed to have special discrete and or
continuous symmetries and to display a quantum phase transition in a
broad sense, that is displaying non-analytical behavior in their
correlation functions.  Naturally these non-analytical behavior can
also be observed in the entanglement properties of pairs of spins in
these models. This route can be followed to develop other models,
e.g models with higher spins or alternating spins on the rungs, with
alternating coupling constants, or next-nearest neighbor
interactions. By exploiting higher dimensional matrices and having
more free parameters at our disposal, we may be able to construct
continuous families of models on spin ladders which have full
rotational symmetry. In this article we have restricted ourselves to
frustration-free Hamiltonians and MPS states with fixed-size.
Relaxing this latter condition allows one to represent any state
$|\psi\ra\in {\mathbb C}^{d\times N}$ as a matrix product state
\cite{vidal, MPSrep} and then one may be able to study more diverse
kinds of phase transitions on spin ladders. These matters will be
taken up in  separate publications.

\section{Acknowledgement} We would like to thank A. Langari for very valuable discussions and the members of the
Quantum information group of Sharif University, specially S. Alipour
and L. Memarzadeh for instructive comments.

{}

\section{Appendix}
In this appendix we briefly discuss the derivation of the explicit
form of the Hamiltonian in terms of local spin operators. For
simplicity we consider models in class A. The Hamiltonian for models
in class B, can be constructed along similar lines. As explained in
the main text, the starting point is to solve the system of
equations
\begin{equation}\label{equations}
    \sum_{i,j,k,l=0}^1 c_{ijkl} A_{ij}A_{kl}=0.
\end{equation}
This solution space is 12 dimensional since we have 4 equations for
16 unknowns. Thus we should find a set of 12 orthogonal vectors
which span this solution space and then form a non-negative linear
combination of the corresponding one dimensional projectors. The
Hamiltonian constructed in this way does not necessarily have the
symmetries imposed on the state, unless we choose new linear
combinations of these vectors which transform suitably under the
symmetry operators. These new basis vectors are found to be:

\begin{figure}\label{labeling}
\centering
    \includegraphics[width=4cm,height=4cm,angle=0]{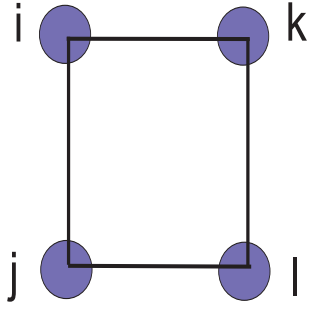}
    \caption{(Color Online) The labeling of the sites in equation (\ref{equations}), for the coefficients $c_{ijkl}$ and the corresponding states $|i\ j\ k\ l\ra$.
    }
\end{figure}
\def\e{\epsilon} \def\s{\sigma}
\begin{eqnarray}
  |2,2\ra &:=&|0000\ra \cr
  |2,1\ra &:=& \frac{1}{2}\left[-\e|0001\ra + |0010\ra -\e |0100\ra + |1000\ra\right] \cr
  |2,0\ra &:=& \frac{1}{\sqrt{6}}\left[-{4a^2}(|0011\ra+|1100\ra) + g\left(|0101\ra +\s\e
  |0110\ra+ \s\e|1001\ra + |1010\ra\right)\right]\cr
 &&\cr
  |1,1\ra &:=& \frac{1}{2}\left[\e|0001\ra + |0010\ra - \e|0100\ra - |1000\ra \right] \cr
  |1,0\ra &:=& \frac{1}{\sqrt{2}}\left[|0110\ra-|1001\ra\right] \cr
&&\cr
  |1',1\ra &:=& \frac{1}{2}\left[-\s\e|0001\ra + |0010\ra +\e |0100\ra -\s |1000\ra\right]  \cr
  |1',0\ra &:=& \frac{1}{\sqrt{2}}\left[|0101\ra-|1010\ra\right] \cr
  &&\cr
  |0,0\ra &:=& \frac{1}{2}\left[|0101\ra -\s\e |0110\ra - \s\e |1001\ra +
  |1010\ra\right], \end{eqnarray}

with $|{\it l},-{\it m}\ra=\sigma_x^{\otimes}|{\it l},{\it m}\ra, \
\ \forall\ \  {\it l}, {\it m}.$\\

Here we have organized the states in multiplets, with a labeling
reminiscent of the one used in labeling the states of $so(3)$
representations. The reason is that in a certain limit
($\epsilon=\sigma=g=-1, a=\frac{1}{2}$) these states actually
comprise the irreducible representations of $so(3)$ which emerge
from the decomposition of the product of four spin one-half
representations living on the sites of two rungs of the ladder. In
fact for the product of these spin 1/2 representations we have
$$(\frac{1}{2})^{\otimes 4}= 2 \oplus 1 \oplus 1'\oplus 1" \oplus 0 \oplus 0'.$$
However only the representations $2, 1, 1'$ and $0$ span the
solution space of (\ref{equations}) in this limit. This labeling is
useful because we can see under what conditions, the Hamiltonian
become fully rotational invariant.  Note also that we have used the
labeling $|i\ j\ k\ l\ra$  in accordance with the labeling of
(\ref{equations}) for $c_{ijkl}$ (see figure (5). The reader can
easily check that each of the above states (or more precisely the
corresponding one dimensional projector) is invariant under the
parity operation $\Pi:\ |i\ j\ k\ l\ra\lo |k\ l\ i\ j\ra$ and
leg-exchange of the ladder $Y:\ |i\ j\ k\ l\ra\lo |j\ i\ l\ k\ra$.
They also transform to each other under the spin-flip operation $X:\
|i\ j\ k\ l\ra\lo |\overline{i}\ \overline{j}\ \overline{k}\
\overline{l}\ra$.

The local Hamiltonian which has the three discrete symmetries
mentioned in the text and the symmetry around rotation along the $z$
axis in spin space, is constructed as follows:
\begin{equation}\label{Ham}
h = \sum_{l=2,1,1',0}\ \sum_{m=0}^l\mu_{lm}(|l,m\ra\la
l,m|+|l,-m\ra\la l,-m|),
\end{equation}
where the coefficients  $\mu_{l,m}$ are non-negative parameters and
together with the parameters $a$ and $g$ form the 10 free parameters
of the Hamiltonian. Of course the total number of coupling constants
(interaction strength) is 8, since we can always shift the ground
state energy and also set the scale of energy by redefintion of
these parameters.  Re-expressing the local operators in terms of
Pauli spin operators and rearranging terms, we find, after a rather
lengthy calculation, the Hamiltonian acting on the ladder where we
have multiplied $h$ by a factor of $8$ for convenience, (see figure
(6) for labeling of sites of the ladder). Note that we use $z_i$ as
an abbreviation to denote the operator $\sigma_{i,z}$ in the
following. The total Hamiltonian is then given by
\def\s{{\pmb{\sigma}}}

\begin{equation}\label{Hamiltonian}
    H = H_1+H_2 +H^{(s)}_{1,2}+H^{z}_{1,2}+H^{(s,z)}_{1,2}
\end{equation}
where
\begin{eqnarray}\label{H1}
    H_1&=&  \sum_i J_0+ J_1 \ z_i z_{i+1} + J_2\
    {\pmb{\sigma}}_i\cdot {\pmb{\sigma}}_{i+1}\cr
H_2&=&  \sum_i  J_0+ J_1 \ z_{i'} z_{i'+1} + J_2\
    {\pmb{\sigma}}_{i'}\cdot {\pmb{\sigma}}_{i'+1},
\end{eqnarray}

\begin{figure}
\centering
    \includegraphics[width=8cm,height=2.5cm,angle=0]{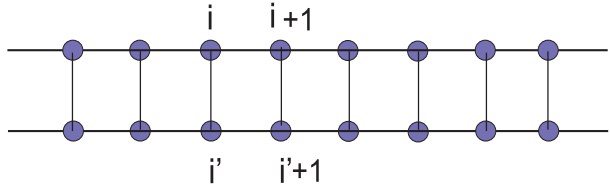}
    \caption{(Color Online) The labeling of sites of the ladder for reading the interaction terms in the Hamiltonian. }
\end{figure}\label{ladder}

\begin{equation}\nonumber
  H^{(z)}_{12} = \sum_i J_8 \ z_i z_{i'} +J_9\ \left(z_i z_{i'+1}+z_{i'}z_{i+1}\right) + J_{10}\  z_i z_{i'} z_{i+1} z_{i'+1}
  ,
\end{equation}

\begin{eqnarray}\nonumber
  H^{(s)}_{12} = \sum_i &J_3&\  \s_i \cdot \s_{i'}+ J_4 \ (\s_i\cdot \s_{i'+1}+\s_{i'}\cdot \s_{i+1})
  +
  J_5\ (\s_i\cdot \s_{i'})(\s_{i+1}\cdot \s_{i'+1}) \cr + &J_6&\  (\s_i\cdot \s_{i+1})(\s_{i'}\cdot
  \s_{i'+1}) + J_7 (\s_i\cdot \s_{i'+1})(\s_{i'}\cdot
  \s_{i+1}) ,
  \end{eqnarray}

and

\begin{eqnarray}\nonumber
  H^{(z,s)}_{12} = \sum_i &J_{11}& \left( z_i z_{i'}(\s_{i+1}\cdot \s_{i'+1}) + z_{i+1} z_{i'+1}(\s_{i}\cdot \s_{i'})\right)  \cr
  +&J_{12}&
\left( z_i z_{i+1}(\s_{i'}\cdot \s_{i'+1}) + z_{i'}
z_{i'+1}(\s_{i}\cdot \s_{i+1})\right) \cr + &J_{13}& \left( z_i
z_{i'+1}(\s_{i}\cdot \s_{i+1}) + z_{i'} z_{i+1}(\s_{i}\cdot
\s_{i'+1})\right).
\end{eqnarray}
 Thus there are bond interaction and plaquette interaction in the
 Hamiltonian all depending on 10 parameters.
The coupling constants are related to these parameters as follows:

\def\si{\sigma}
\begin{eqnarray}\label{coupling}\nonumber
J_0&=&\m_{22}+4(\m_{21}+\m_{11}+\m_{1'1})+\mu_{10}-\m_{1'0}+2(\mu_{00}+\mu_{20})+16a^4
    \mu_{20}\cr
J_1 &=&
\m_{22}+\frac{1}{2}(-\m_{21}+\m_{11}+\si\m_{1'1}+\m_{1'0}-\m_{10})+\frac{4}{3}a^2(g\si\e-2a^2)\m_{20}\cr
J_2&=&\frac{1}{2}(\m_{21}-\m_{11}-\si\m_{1',1})-\frac{4}{3}\si\e a^2
g\m_{20}\cr
J_3&=&-\e(\m_{21}-\m_{11})-\si\e\m_{1'1}+\e\si(\frac{2}{3}g^2\m_{20}-\m_{00})\cr
J_4 &=&-\frac{1}{2}\e(\m_{21}+\m_{11}-\m_{1'1})-\frac{4}{3}a^2
g\m_{20}\cr J_{5}
&=&\frac{1}{2}(\m_{00}-\m_{10}-\m_{1'0})+\frac{1}{3}\m_{20}(g^2-8a^4)\cr
J_{6} &=& \frac{8}{3}a^4\m_{20}+ \frac{1}{2}(\m_{1'0}-\m_{10})\cr
J_{7}&=& \frac{8}{3}a^4\m_{20}+\frac{1}{2}(\m_{10}-\m_{1'0})\cr J_8
&=& 2\m_{22}+\e(\m_{21}-\m_{11})+\si\e
\m_{1'1}-\m_{10}-\m_{1'0}+(\si\e-1)\m_{00}+\frac{2}{3}\m_{20}(8a^4-(1+\si\e)g^2)\cr
J_{9}&=&
\m_{22}+\frac{1}{2}\e(\m_{21}+\m_{11}-\m_{1'1})+\frac{1}{2}(\m_{10}-\m_{1'0})+\frac{4}{3}a^2(g-2a^2)\m_{20}\cr
J_{10}&=&\m_{22}+2(\e-1)\m_{21}+(\e+1)(\si-1)\m_{1'1}+(1-\si\e)\m_{00}+\frac{2}{3}(1+\si\e)g^2\m_{20}+\frac{8}{3}a^2(2a^2-g-g\si\e)\m_{20}\cr
J_{11} &=& \frac{1}{2}(-\e\m_{21}+\e\m_{11}-\e\si
\m_{1'1}+\mu_{10}+\mu_{1'0})+\frac{1}{2}(\e\si-1)\m_{00}+\frac{\mu_{20}}{3}(8a^4-(1+\si\e)g^2)\cr
J_{12} &=&
\frac{1}{2}(\m_{21}-\m_{11}-\si\m_{1'1}+\m_{10}-\m_{1'0})-\frac{4}{3}a^2(2a^2-g\si\e)\m_{20}\cr
J_{13} &=&
\frac{1}{2}(-\e\m_{21}-\e\m_{11}+\e\m_{1'1}-\m_{10}+\m_{1'0})+\frac{4}{3}a^2\m_{20}(g-2a^2).
\end{eqnarray}
This is a complicated looking Hamiltonian, with many types of
interactions, but in view of the large number of parameters, it is
possible to look at specific subsets of the parameter space, where
some of the interactions are absent. Indeed the parameter space
consists of four disconnected parts, each of which corresponds to
one choice of the pair $(\sigma, \epsilon)$. Let us for example
consider the subset on which the Hamiltonian has full rotational
symmetry.
 It is well known that any operator of the form
$\sum_{m=-l}^l|l,m\ra\la l,m|$ is a scalar. Thus in the limit
$g=\epsilon=\si=-1, a=\frac{1}{2}$ if we set $\mu_{2m}=:6\mu, \
\m_{1,m}=:2\nu, \ \m_{1',m}=:2\xi \ \ \ \forall \ \ m$ and
$\m_{0,0}=:2\eta$, then we will have a Hamiltonian which is fully
rotational invariant. On can see that in this limit, all the
coupling constant corresponding to non-scalar terms in the
Hamiltonian vanish and we are left with
\begin{eqnarray}\nonumber
J_0&=& 48\mu+10\nu+6\xi+4\mu\cr
  J_2 &=& 5\mu-\nu+\xi \cr
  J_3 &=& 10\m - 2\n -2\xi - 2\eta \cr
  J_4 &=& 5\mu+\n-\xi \cr
  J_5 &=& \mu-\nu-\xi+\eta \cr
  J_6 &=& \mu+\xi-\n\cr
  J_7&=& \m+\n-\xi.
\end{eqnarray}

\end{document}